\documentstyle[prl,aps,preprint,psfig]{revtex}
\tighten
\begin{document}

\title{Study of the $P$-wave charmonium state $\chi_{cJ}$ in $\psi(2S)$ decays}
\author{
J.~Z.~Bai$^{1}$,   J.~G.~Bian$^{1}$,
I.~Blum$^{11}$,
Z.~W.~Chai$^{1}$,  G.~P.~Chen$^{1}$,
H.~F.~Chen$^{10}$,
J.~Chen$^3$,
J.~C.~Chen$^{1}$,  Y.~Chen$^{1}$,    Y.~B.~Chen$^{1}$,
Y.~Q.~Chen$^{1}$,  B.~S.~Cheng$^{1}$  X.~Z.~Cui$^{1}$,   H.~L.~Ding$^{1}$,
L.~Y.~Ding$^{1}$,  L.~Y.~Dong$^{1}$,  Z.~Z.~Du$^{1}$,
W.~Dunwoodie$^7$,
S.~Feng$^{1}$,
C.~S.~Gao$^{1}$,
M.~L.~Gao$^{1}$,   S.~Q.~Gao$^{1}$,
P.~Gratton$^{11}$,
J.~H.~Gu$^{1}$,    S.~D.~Gu$^{1}$,
W.~X.~Gu$^{1}$,    Y.~F.~Gu$^{1}$,    Y.~N.~Guo$^{1}$,   S.~W.~Han$^{1}$,
Y.~Han$^{1}$,
F.~A.~Harris$^8$,
J.~He$^{1}$,       J.~T.~He$^{1}$,    M.~He$^{5}$,
D.~G.~Hitlin$^2$,
G.~Y.~Hu$^{1}$,    H.~M.~Hu$^{1}$,    J.~L.~Hu$^{12,1}$,    Q.~H.~Hu$^{1}$,
T.~Hu$^{1}$,       X.~Q.~Hu$^{1}$,    J.~D.~Huang$^{1}$, Y.~Z.~Huang$^{1}$,
J.~M.~Izen$^{11}$,
C.~H.~Jiang$^{1}$, Y.~Jin$^{1}$,      Z.~J.~Ke$^{1}$,
M.~H.~Kelsey$^2$,  B.~K.~Kim$^{11}$,  D.~Kong$^8$,
Y.~F.~Lai$^{1}$,
P.~F.~Lang$^{1}$,
A.~Lankford$^{9}$,
C.~G.~Li$^{1}$,    D.~Li$^{1}$,       H.~B.~Li$^{1}$,
J.~Li$^{1}$,       P.~Q.~Li$^{1}$,    R.~B.~Li$^{1}$,    W.~Li$^{1}$,
W.~D.~Li$^{1}$,    W.~G.~Li$^{1}$,    X.~H.~Li$^{1}$,    X.~N.~Li$^{1}$,
H.~M.~Liu$^{1}$,   J.~Liu$^{1}$,      J.~H.~Liu$^{1}$,   R.~G.~Liu$^{1}$,
Y.~Liu$^{1}$,
X.~C.~Lou$^{11}$,  B.~Lowery$^{11}$,
F.~Lu$^{1}$,       J.~G.~Lu$^{1}$,    J.~Y.~Lu$^{1}$,
L.~C.~Lu$^{1}$,    C.~H.~Luo$^{1}$,   A~.M~.Ma$^{1}$,    E.~C.~Ma$^{1}$,
J.~M.~Ma$^{1}$,
R.~Malchow$^3$,
H.~S.~Mao$^{1}$,   Z.~P.~Mao$^{1}$,   X.~C.~Meng$^{1}$,
J.~Nie$^{1}$,
S.~L.~Olsen$^8$,   J.~Oyang$^2$,      D.~Paluselli$^8$, L.~J.~Pan$^8$,
J.~Panetta$^2$,    F.~Porter$^2$,
N.~D.~Qi$^{1}$,    X.~R.~Qi$^{1}$,    C.~D.~Qian$^{6}$,
J.~F.~Qiu$^{1}$,   Y.~H.~Qu$^{1}$,    Y.~K.~Que$^{1}$,   G.~Rong$^{1}$,
M.~Schernau$^9$,
Y.~Y.~Shao$^{1}$,  B.~W.~Shen$^{1}$,  D.~L.~Shen$^{1}$,  H.~Shen$^{1}$,
X.~Y.~Shen$^{1}$,  H.~Y.~Sheng$^{1}$, H.~Z.~Shi$^{1}$,   X.~F.~Song$^{1}$,
J.~Standifird$^{11}$,
F.~Sun$^{1}$,      H.~S.~Sun$^{1}$,   S.~Q.~Tang$^{1}$,
W.~Toki$^3$,
G.~L.~Tong$^{1}$,  G.~Varner$^{8}$,  
F.~Wang$^{1}$,     L.~S.~Wang$^{1}$,  L.~Z.~Wang$^{1}$,  M.~Wang$^{1}$,
Meng~Wang$^{1}$,   P.~Wang$^{1}$,     P.~L.~Wang$^{1}$,  S.~M.~Wang$^{1}$,
T.~J.~Wang$^{1}$\cite{atNU0},  Y.~Y.~Wang$^{1}$,  
M.~Weaver$^2$,
C.~L.~Wei$^{1}$,   Y.~G.~Wu$^{1}$,
D.~M.~Xi$^{1}$,    X.~M.~Xia$^{1}$,   P.~P.~Xie$^{1}$,   Y.~Xie$^{1}$,
Y.~H.~Xie$^{1}$,   W.~J.~Xiong$^{1}$, C.~C.~Xu$^{1}$,    G.~F.~Xu$^{1}$,
S.~T.~Xue$^{1}$,   J.~Yan$^{1}$,      W.~G.~Yan$^{1}$,
C.~M.~Yang$^{1}$,  C.~Y.~Yang$^{1}$,  J.~Yang$^{1}$,     
W.~Yang$^3$,
X.~F.~Yang$^{1}$,
M.~H.~Ye$^{1}$,    S.~W.~Ye$^{10}$,    Y.~X.~Ye$^{10}$,    K.~Yi~$^{1}$,
C.~S.~Yu$^{1}$,    C.~X.~Yu$^{1}$,    Y.~H.~Yu$^{4}$,   Z.~Q.~Yu$^{1}$,
Z.~T.~Yu$^{1}$,    C.~Z.~Yuan$^{12,1}$,  Y.~Yuan$^{1}$,     B.~Y.~Zhang$^{1}$,
C.~C.~Zhang$^{1}$, D.~H.~Zhang$^{1}$, Dehong ~Zhang$^{1}$,
H.~L.~Zhang$^{1}$, J.~Zhang$^{1}$,    J.~L.~Zhang$^{1}$, J.~W.~Zhang$^{1}$,
L.~S.~Zhang$^{1}$, Q.~J.~Zhang$^{1}$, S.~Q.~Zhang$^{1}$, X.~Y.~Zhang$^{5}$,
Y.~Zhang$^{1}$,    Y.~Y.~Zhang$^{1}$, D.~X.~Zhao$^{1}$,  
H.~W.~Zhao$^{1}$,
J.~W.~Zhao$^{1}$,  M.~Zhao$^{1}$,    W.~R.~Zhao$^{1}$, Z.~G.~Zhao$^{1}$,
 J.~P.~Zheng$^{1}$,
L.~S.~Zheng$^{1}$, Z.~P.~Zheng$^{1}$, G.~P.~Zhou$^{1}$,  H.~S.~Zhou$^{1}$,
L.~Zhou$^{1}$,     Q.~M.~Zhu$^{1}$,   Y.~C.~Zhu$^{1}$,   Y.~S.~Zhu$^{1}$,
B.~A.~Zhuang$^{1}$
\\ (BES Collaboration)}
\address{
$^1$Institute of High Energy Physics, Beijing 100039, People's Republic of
 China\\
$^2$California Institute of Technology, Pasadena, California 91125\\
$^3$Colorado State University, Fort Collins, Colorado 80523\\
$^4$Hangzhou University, Hangzhou 310028,
People's Republic of China\\
$^5$Shandong University, Jinan 250100, People's Republic of
 China\\
$^6$Shanghai Jiaotong University, Shanghai 200030,
People's Republic of China\\
$^7$Stanford Linear Accelerator Center, Stanford, California 94309\\
$^8$University of Hawaii, Honolulu, Hawaii 96822\\
$^9$University of California at Irvine, Irvine, California 92717\\
$^{10}$University of Science and Technology of China, Hefei 230026,
People's Republic of China\\
$^{11}$University of Texas at Dallas, Richardson, Texas 75083-0688\\
$^{12}$China Center of Advanced Science and Technology (World Laboratory), 
Beijing 100080, People's Republic of China}
\date{Received 1 July 1998}
\maketitle

\begin{abstract}
The processes $\psi(2S)\rightarrow \gamma \pi^+ \pi^-$,
$\gamma K^+ K^-$ and $\gamma p \bar{p}$ have been studied
using a sample of $3.79 \times 10^6$ $\psi(2S)$
decays. We determine the total width of the $\chi_{c0}$ to be
$\Gamma^{tot}_{\chi_{c0}} = 14.3\pm 2.0\pm 3.0$~MeV.
We present the first measurement of the branching fraction
$B(\chi_{c0} \rightarrow p \bar{p}) =
(15.9 \pm 4.3 \pm 5.3)\times 10^{-5}$, where
the first error is statistical and the second one systematic.
Branching fractions of $\chi_{c0,2} \rightarrow \pi^+ \pi^-$
and $K^+ K^-$ are also reported.
\end{abstract}


%
The hadronic decay rates of the $P$-wave quarkonium states provide
tests of perturbative QCD.  Recently, a systematic approach
to the treatment of the infrared ambiguities in the calculation
of the production and decays of these states has been developed~\cite{BBL}.
However, existing experimental information on the
triplet $P$-wave $c\bar{c}$ states ($\chi_{c0,1,2}$), especially
the $J=0$ $\chi_{c0}$, is not adequate for testing the
predictions of this new theory.

In particular, the total width of the $\chi_{c0}$ is a quantity
of considerable interest.  The two existing measurements
have large errors and only marginal consistency~\cite{cbal}.  Also
of interest is the decay $\chi_{c0}\rightarrow p\bar{p}$, which
is forbidden in the limit of massless helicity conservation~\cite{brodsky}
and has been calculated by many different models~\cite{masscor,quark-di}.
Here the only existing measurement is an upper limit on the partial width
that does not seriously constrain the theory~\cite{brandelik}.  
Calculations of the branching fractions for other exclusive 
$\chi_{cJ}$ decays, such as $\chi_{cJ}\rightarrow 
\pi^+\pi^-$~\cite{chernyak}, has revealed orders-of-magnitude discrepancies
with the data reported by early experiments.  For these reasons, 
measurements of these properties of the $\chi_{cJ}$ states with 
improved precision are very useful.

In this paper we report a measurement of 
$\Gamma^{tot}_{\chi_{c0}}$ determined from an analysis
of exclusive $\psi(2S)\rightarrow \gamma\pi^+\pi^-$ and $\gamma K^+K^-$
decays seen in the Beijing Spectrometer (BES) at the
Beijing Electron Positron Collider (BEPC).  
We also report a first measurement of the $\chi_{c0}\rightarrow p\bar{p}$
branching fraction, and improved precision on the branching fractions for 
$\chi_{c0,2}\rightarrow \pi^+\pi^-$ and $K^+K^-$.

%
The BES is a conventional solenoidal magnet detector that is
described in detail in Ref.~\cite{bes}. A four-layer central
drift chamber (CDC) surrounding the beampipe provides trigger
information. A forty-layer main drift chamber (MDC), located
radially outside the CDC, provides trajectory and energy loss
($dE/dx$) information for charged tracks over $85\%$ of the
total solid angle.  The momentum resolution is
$\sigma _p/p = 0.017 \sqrt{1+p^2}$ ($p$ in $\hbox{\rm GeV}/c$),
and the $dE/dx$ resolution for hadron tracks is $\sim 11\%$.
An array of 48 scintillation counters surrounding the MDC  measures
the time of flight (TOF) of charged tracks with a resolution of
$\sim 450$ ps for hadrons.  Radially outside the TOF system is a 12
radiation length, lead-gas barrel shower counter (BSC).  This
measures the energies
of electrons and photons over $\sim 80\%$ of the total solid
angle with an energy resolution of $\sigma_E/E=22\%/\sqrt{E}$ ($E$
in GeV).  Outside of the solenoidal coil, which
provides a 0.4~T magnetic field over the tracking volume,
is an iron flux return that is instrumented with
three double layers of  counters that
identify muons of momentum greater than 0.5~GeV/c.

We study $\chi_{c}$ states produced by the reaction
$e^+e^-\rightarrow\psi(2S)\rightarrow\gamma\chi_{c}$ in a
data sample corresponding to a total of $3.79 \times 10^{6}$
$\psi(2S)$ decays~\cite{npsp}.
For the $\Gamma^{tot}_{\chi_{c0}}$ determination reported here
we use the paired pseudoscalar
meson decay modes of $\chi_{c0,2}\rightarrow\pi^+ \pi^-$ and $K^+ K^-$.
Using the particle identification
capabilities of the detector and four-constraint kinematic fits,
we can get relatively pure event samples.
Moreover, since the decays of the $\chi_{c1}$
and the $\eta_{c}^\prime$ to $\pi^+\pi^-$ or $K^+K^-$ are forbidden by
parity conservation, the $\chi_{c0}$ and $\chi_{c2}$ signals in
these channels are free of distortions due to possible contamination
of these other states.
The effects of cross-contamination between the $\pi^+\pi^-$
and $K^+K^-$ event samples are
estimated by Monte Carlo (MC) simulations and corrected accordingly.
The total width of the $\chi_{c2}$ has been precisely
measured to be
$\Gamma^{tot}_{\chi_{c2}} = 2.00\pm 0.18$~MeV~\cite{pdg},
which is much narrower than our experimental resolution
at $M_{\chi_{c2}}$ of 7.83~MeV.
Thus, the strong $\chi_{c2}\rightarrow\pi^+\pi^-$
signal in our data is used to provide a
direct experimental determination of our resolution.
We only rely on the MC simulation to determine how the
resolution changes between $M_{\chi_{c2}}$
and $M_{\chi_{c0}}$.

We select  $\psi(2S)\rightarrow \gamma \pi^+\pi^-$, $\gamma K^+K^-$ and
$\gamma p \bar{p}$ by imposing the following selection
criteria.

    A cluster of deposited energy in the BSC
is regarded as a photon candidate if:
    (1) the angle between the nearest charged track and the cluster in 
the $r\phi$ plane is greater than $15^{\circ}$; 
    (2) the energy of the cluster is greater than 20~MeV and some energy 
is deposited in the first 6 radiation lengths of the counter; and \
    (3) the angle determined from the cluster development in the
BSC  agrees with that determined from the relative position
of the shower location and the interaction point to within
$37^{\circ}$. 
    At least one and at most three photon candidates are
allowed in an event. The candidate with the largest
BSC energy is assumed to be the photon radiated from
the $\psi(2S)$.

In addition, we require that the event has two 
oppositely signed charged tracks in the MDC that
both have at least 13 good hits and are well fit to 
a three-dimensional helix. Events with tracks where 
the $dE/dx$ measured in the MDC and the shower properties
in the BSC are consistent with electrons are rejected.
For each track, the TOF and $dE/dx$ measurements are used to assign 
probabilities that the particle is
a pion, kaon and proton ($Prob_{\pi},Prob_{K},Prob_{p}$).
We require both tracks to
have $Prob_{\pi}>0.01$ (for $\pi^+ \pi^-$) or
$Prob_{K}>0.01$ (for $K^+ K^-$) or $Prob_{p}>0.05$ (for $p\bar{p}$).
In addition we do four-constraint kinematic fits to the hypotheses
$\psi(2S) \rightarrow \gamma \pi^+ \pi^-$,
$\psi(2S) \rightarrow \gamma K^+ K^-$ and
$\psi(2S) \rightarrow \gamma p\bar{p}$ and require
the $\chi^2$ probability of
the fit $P_{\chi^{2}}$ to be greater than 0.01 for $\pi^+ \pi^-$
or $K^+ K^-$ and greater than 0.05 for $p\bar{p}$.

There is some background from $J/\psi\rightarrow\mu^+\mu^-$,
where the $J/\psi$ is produced by cascade $\psi(2S)$ to $J/\psi$ decays.
To reduce this, we reject events
where the response of the muon detection
system is consistent with the two charged tracks being muons.
The surviving $\mu^+\mu^-$ background events do not 
populate the $\pi^+\pi^-$ invariant mass distribution 
near the $\chi_{c0}$ or $\chi_{c2}$ masses.  
In the $K^+K^-$ mass distribution, however, they 
populate the region in the lower mass side of $\chi_{c0}$, 
and cause an abnormal distribution.  
For the $p\bar{p}$ sample, the $\mu^+\mu^-$ background 
level is significant.  For this channel,
in order to insure that both tracks are directed at the sensitive
area of the muon detection system, we require
$|\cos\theta_{\hbox{MDC}}|<0.65$ for both the $p$ and the 
$\bar{p}$ track.

    To distinguish $\gamma \pi^+ \pi^-$ from $\gamma K^+ K^-$,
we define
\[ Prob^{P^+P^-}_{all}=Prob(\chi^{2}_{\hbox{all}},
      \hbox{\em ndf}_{\hbox{all}}), \]
where 
\( \chi^{2}_{\hbox{all}} = 
\chi^{2}_{4C}+\chi^{2}_{TOF}+\chi^{2}_{dE/dx} \) and 
\( \hbox{\em ndf}_{\hbox{all}} =
   \hbox{\em ndf}_{4C}+\hbox{\em ndf}_{TOF}+\hbox{\em ndf}_{dE/dx} \)
are the total $\chi^2$ and 
the corresponding number of degrees of freedom of
the $\chi^2$ distribution. Here
$\chi^{2}_{4C}$, $\chi^{2}_{TOF}$ and $\chi^{2}_{dE/dx}$ correspond
to the $\chi^2$ values from the
4-constraint kinematic fit, the TOF measurements for the $\pi$
or $K$ hypothesis, and  the $dE/dx$ measurements for 
the $\pi$ or $K$ hypothesis, respectively, and
$\hbox{\em ndf}_{4C}$, $\hbox{\em ndf}_{TOF}$ and $\hbox{\em ndf}_{dE/dx}$
are the corresponding numbers of degrees of freedom.
If $Prob^{\pi^+ \pi^-}_{all} > Prob^{K^+ K^-}_{all}$,
the event is categorized as a $\gamma \pi^+ \pi^-$ event,
and if $Prob^{K^+ K^-}_{all} > Prob^{\pi^+ \pi^-}_{all}$, it
is categorized as a $\gamma K^+ K^-$ event.

\begin{figure}
\centerline{\hbox{
\psfig{file=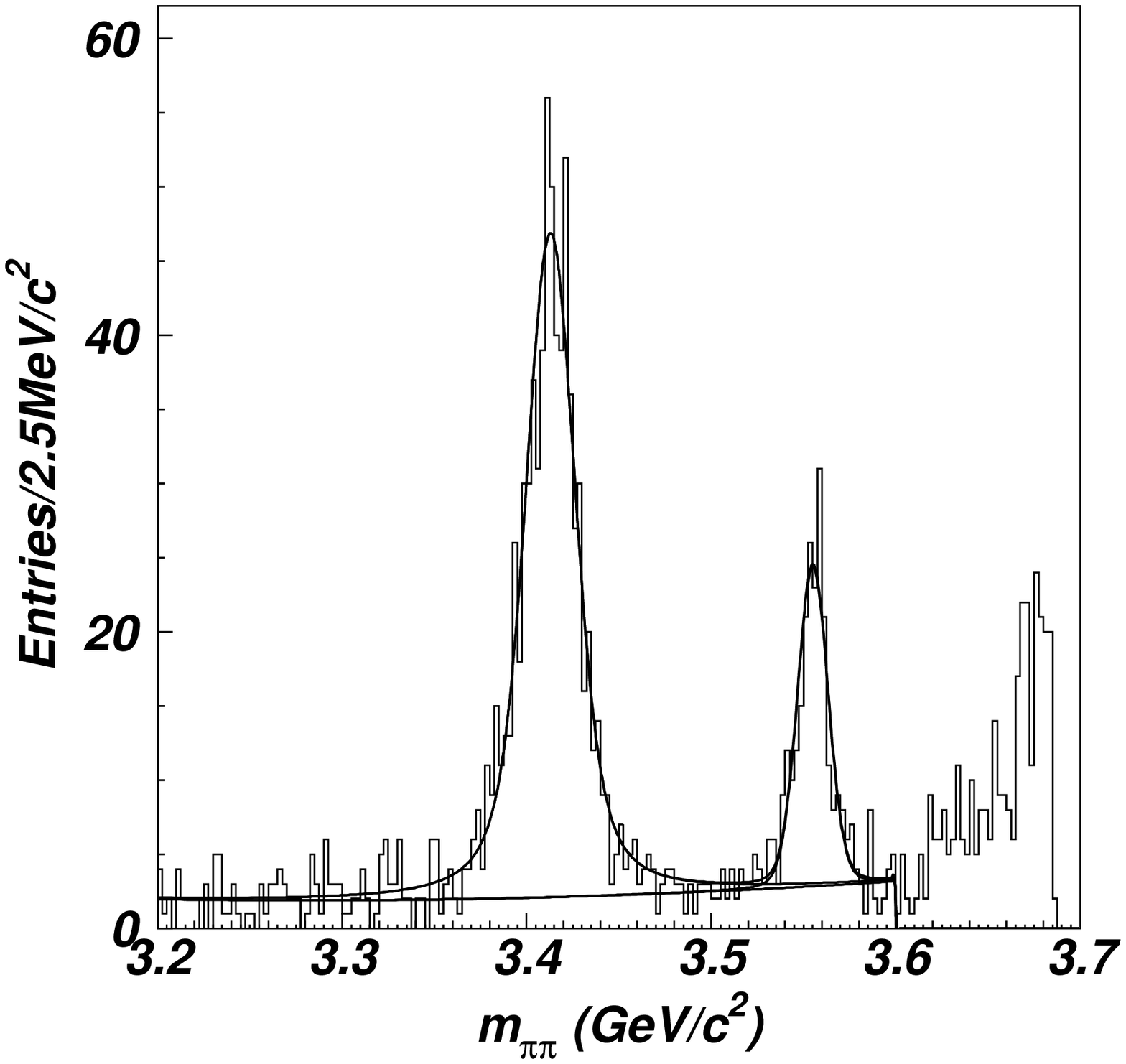,height=5.5cm}}}
\caption{The $\pi^+ \pi^-$ mass distribution for selected
$\psi(2S) \rightarrow \gamma \pi^+ \pi^-$ events.
\label{chic02pi_unbin_fig}}
\end{figure}

\begin{figure}
\centerline{\hbox{
\psfig{file=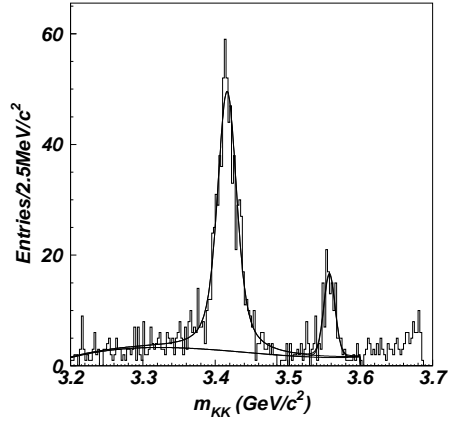,height=5.5cm}}}
\caption{The $K^+ K^-$ mass distribution for selected
$\psi(2S) \rightarrow \gamma K^+ K^-$ events. }
\label{chic02ka_unbin_fig}
\end{figure}

\begin{figure}
\centerline{\hbox{
\psfig{file=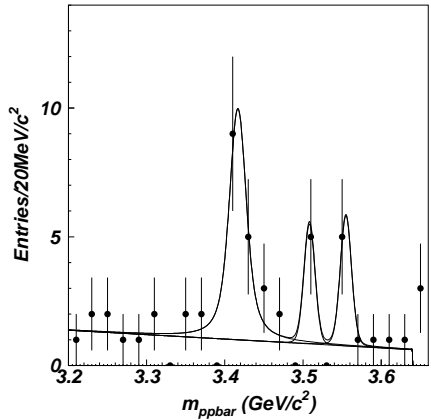,height=5.5cm}}}
\caption{The $p \bar{p}$ mass distribution for selected
$\psi(2S) \rightarrow \gamma p \bar{p}$ events.
\label{chic012_unbin}}
\end{figure}

Figures~\ref{chic02pi_unbin_fig} and \ref{chic02ka_unbin_fig}
show the $\pi^+ \pi^-$ and $K^+ K^-$ invariant mass distributions
after the imposition of all the above-listed selection requirements.
In these plots,  the mass values corresponding to the
$\chi_{c0}$ and $\chi_{c2}$ peaks are
lower in the  $\pi^+ \pi^-$ channel and higher for
$K^+ K^-$, indicating the presence of some remaining cross contamination
between the two samples, which must be accounted for
in the determination of the $\chi_{c0}$
parameters.  From Fig.~\ref{chic02ka_unbin_fig}, it is apparent that
the $\chi_{c2} \rightarrow K^+ K^-$ sample is
statistically limited.  We therefore use only 
the $\chi_{c2}\rightarrow\pi^+\pi^-$ signal
to calibrate the mass resolution.
In Fig.~\ref{chic012_unbin}, the $p\bar{p}$ invariant
mass distribution, there is a clear $\chi_{c0}$
signal and evidence for the $\chi_{c1}$ and  $\chi_{c2}$.

%
     We use Monte Carlo simulated data to determine the
$\pi^+ \pi^-$ and $K^+ K^-$ cross contamination
probabilities, the detection efficiencies and the mass
resolutions.  We generate events assuming that the reaction
$\psi(2S)\rightarrow\gamma\chi_{cJ}$ is a pure E1
transition.  The decays
$\chi_{cJ}\rightarrow$~pseudoscalar meson pairs and
$\chi_{c0} \rightarrow p \bar{p}$
have only one independent helicity
amplitude and are thus unambiguous~\cite{generator}.  For
$\chi_{c1,2} \rightarrow p \bar{p}$ decays, there are
no available experimental data on the
helicity amplitudes, and we use an isotropic
distribution.  (Our $p\bar{p}$ event samples are too small to
permit a helicity amplitude analysis.)

We subject the MC-generated events to the same
selection process as is used for the data and determine the
detection efficiencies for each mode.
For the $p\bar{p}$ mode, the detection efficiencies
and the error caused by the limited statistics of the
Monte Carlo sample are
$\varepsilon_{\chi_{c0}}= (27.1 \pm 0.6)\%$,
$\varepsilon_{\chi_{c1}}= (30.3 \pm 0.7)\%$ and
$\varepsilon_{\chi_{c2}}= (27.6 \pm 0.6)\%$,  and mass resolutions at
the $\chi_{c0}$, $\chi_{c1}$ and $\chi_{c2}$ equal to 7.3,
6.8 and 6.7~MeV, respectively.
       For the $\pi^+\pi^-$ and $K^+K^-$ modes, the
simulation shows that  the mass dependence of the
detection efficiency is small and the mass resolution
function is very nearly Gaussian.  We compensate
for the distortion of the mass
spectra due to the the $\pi^+\pi^-$-$K^+K^-$
cross contamination by
calibrating the mass resolution derived from the MC simulation with
the $\chi_{c2}\rightarrow \pi^+\pi^-$ line shape seen with
the data.  The efficiencies are
$\varepsilon_{\chi_{c0} \rightarrow \pi^+\pi^-} = (36.9 \pm 0.3)\%$,
$\varepsilon_{\chi_{c2} \rightarrow \pi^+\pi^-} = (38.9 \pm 0.5)\%$,
$\varepsilon_{\chi_{c0} \rightarrow K^+K^-} = (32.8 \pm 0.3)\%$ and
$\varepsilon_{\chi_{c2} \rightarrow K^+K^-} = (34.9 \pm 0.5)\%$, and
the probability for $\chi_{c0}~(\chi_{c2})\rightarrow K^+K^-$ events to be
categorized as $\pi^+\pi^-$ is $(5.6 \pm 0.2)\%$
($(6.0 \pm 0.2)\%$), and that for
$\chi_{c0}~(\chi_{c2})\rightarrow\pi^+\pi^-$ events to be selected as
$K^+K^-$ is $(7.1  \pm 0.2)\%$  ($(7.4 \pm 0.3)\%$), where the error 
is from the statistics of the Monte Carlo sample.

%
      The invariant mass distributions in 
Figs.~\ref{chic02pi_unbin_fig},~\ref{chic02ka_unbin_fig} and 
\ref{chic012_unbin} are fit by using an unbinned 
maximum likelihood algorithm. 
      For the $p \bar{p}$ channel, the invariant
mass region between 3.20 and 3.64~GeV is fit
with three Breit-Wigner resonances 
plus a linear background function.
The Breit-Wigner resonance width for the 
$\chi_{c0}$ is fixed at $14.3$~MeV, the value determined
from an analysis of $\chi_{c0}\rightarrow\pi^+\pi^-$ decays described
below; those for the $\chi_{c1}$  and
$\chi_{c2}$ are fixed at the PDG values~\cite{pdg}.  
The resonances are  smeared by Gaussian functions
with rms widths fixed at the MC-determined
mass resolution values.  The fit result, shown as the
curve in Fig.~\ref{chic012_unbin}, gives
$15.2\pm 4.1, 4.2\pm 2.2$ and $4.7\pm 2.5$
events for the $\chi_{c0},\chi_{c1}$
and $\chi_{c2}$ states, respectively.

      For the $\pi^+ \pi^-$ channel, we first fit the invariant 
mass region between 3.5 and 3.6~GeV with a Breit-Wigner resonance
with $\Gamma_{\chi_{c2}}$ fixed at the PDG value of $2.00$~MeV, smeared 
by a Gaussian resolution function with an rms width that is
allowed to float.  We also include
a linear background function.  The fit results in a
$\chi_{c2}$ mass resolution 
of $7.83 \pm 1.04$~MeV, which is slightly higher than the 
MC result of $6.31 \pm 0.11$~MeV.
We scale the  MC value for the mass resolution
at the $\chi_{c0}$  ($8.12 \pm 0.23$~MeV) 
by the ratio of the fitted MC results at
the $\chi_{c2}$ and get a
mass resolution  at the $\chi_{c0}$ of $10.08$~MeV.

     We then fit the  $\pi^+ \pi^-$ mass spectrum between 3.2 and
3.6~GeV with two Gaussian-smeared Breit-Wigners with
resolutions fixed at $10.08$ and $7.83$~MeV, and a second order polynomial
background function, and with $\Gamma_{\chi_{c2}}$ fixed
at the PDG value of $2.00$~MeV.
The fit, shown as the curve in Fig.\ \ref{chic02pi_unbin_fig}, gives
$ \Gamma_{\chi_{c0}} = 14.3 \pm 2.0 $~MeV,
where the error is statistical.
The fitted numbers of $\chi_{cJ} \rightarrow \pi^+ \pi^-$ events
are
\(n^{obs}_{\chi_{c0}\rightarrow \pi^+ \pi^-} =  720 \pm 32 \) and
\(n^{obs}_{\chi_{c2}\rightarrow \pi^+ \pi^-} =  185 \pm 16 \),
where the errors are statistical.

    We fit the  $K^+ K^-$  mass spectrum between 3.2
and 3.6~GeV to two 
Gaussian-smeared Breit-Wigner resonance functions plus a 
background function that includes the possibility of
distortions to the line shape due to $\mu^+\mu^-X$ background
events.  (Because this mode is not used for width determination, 
a high precision knowledge of the mass resolution
is not an issue.) The $\chi_{c2}$ width is fixed.
The resulting fit, shown in Fig.~\ref{chic02ka_unbin_fig}, gives
\(n^{obs}_{\chi_{c0}\rightarrow K^+ K^- } = 774 \pm 38 \)  and
\(n^{obs}_{\chi_{c2}\rightarrow K^+ K^- } = 115 \pm 13, \)
where the errors are statistical.

%
Errors in the determination of $\Gamma_{\chi_{c0}}$ are 
caused by the uncertainty in $\Gamma_{\chi_{c2}}$, the 
determination of the mass resolution, the shape of the 
background, the mass dependence of the efficiency correction,
and the choice of experimental cuts.   We 
add the estimated errors from these
sources in quadrature and get a total relative 
systematic error on $\Gamma_{\chi_{c0}}$ of 21\%.

Systematic errors on the branching fractions, which arise
from the uncertainties in $\Gamma_{\chi_{c2}}$,
the mass resolution, the choice of the background function,
the efficiency determination, and  the choices of the selection criteria
are 11.5\%, 12.6\%, 12.1\% and 14.7\%
for $B(\chi_{c0} \rightarrow \pi^+ \pi^-)$,
$B(\chi_{c0} \rightarrow K^+ K^-)$,
$B(\chi_{c2} \rightarrow \pi^+ \pi^-)$
and $B(\chi_{c2} \rightarrow K^+ K^-)$, respectively.
There are overall errors
caused by the uncertainty of the total number of $\psi(2S)$ events
and the uncertainties in the
$\psi(2S) \rightarrow \gamma \chi_{cJ}$ branching fractions.
Adding these errors in quadrature gives  total relative 
systematic errors of 14\%, 15\%, 15\% and 16\%, respectively,  for
$B(\chi_{c0} \rightarrow \pi^+ \pi^-)$,
$B(\chi_{c0} \rightarrow K^+ K^-)$,
$B(\chi_{c2} \rightarrow \pi^+ \pi^-)$
and $B(\chi_{c2} \rightarrow K^+ K^-)$.

       For $B(\chi_{cJ} \rightarrow p \bar{p})$,
sources of systematic errors
include those listed above plus that associated with the
assumption of an isotropic
angular distribution for $\chi_{c1,2}\rightarrow p\bar{p}$ decays. 
Adding all of the errors in quadrature gives relative systematic 
errors of 33\%, 67\% and 56\% for the $\chi_{c0}$, 
$\chi_{c1}$ and $\chi_{c2}$ states, respectively.

%
In summary, we obtain the total width of the $\chi_{c0}$ to be
\[\Gamma_{\chi_{c0}} = 14.3\pm 2.0\pm 3.0 \, \, \hbox{\rm MeV}, \]
where the first error is statistical and the second is
systematic. The $\chi_{cJ}$ branching fraction results 
are listed in Table\ \ref{pika} and Table\ \ref{ppb}.
The measured width for the $\chi_{c0}$ is consistent
with but substantially more precise than the previous 
measurement~\cite{cbal} (the uncertainty is reduced from ~40\% to ~25\%).
The calculations involving new factorization schemes 
with high order QCD corrections~\cite{huang,andrea} are in good 
agreement with our measurement.

Our branching fraction for $\chi_{c0}\rightarrow p \bar{p}$ is
the first measurement for this decay, and is compatible with the
previous upper bound~\cite{brandelik}. Our results for
$\chi_{c1,2}\rightarrow p \bar{p}$ decays, although statistically
limited, are consistent, within errors, with the
values determined from studies of charmonium states
formed directly in $p \bar{p}$ annihilation~\cite{e760r704}.
The calculation with mass correction effect~\cite{masscor}
gives much smaller value of $\Gamma(\chi_{c0}\rightarrow p \bar{p})$
than our result while the model considering the diquark content
of the proton~\cite{quark-di} can find result consistent with 
our measurement.

Finally, our branching fractions for $\chi_{c0,2}$ decays into
$\pi^+ \pi^-$ and $K^+ K^-$ are somewhat lower than the existing world
average~\cite{pdg}. Recent calculations of exclusive $\chi_{cJ}$ 
decays that include contributions from color-octet 
processes~\cite{bolz} are in generally good agreement with 
our measurements. Using our results and canceling out the 
common errors in the branching fractions,  we get the ratios 
of the branching fractions of
\( \frac{B(\chi_{c0} \rightarrow \pi^+ \pi^-)}
        {B(\chi_{c0} \rightarrow K^+ K^-    )}
    = 0.82 \pm 0.15    \) and
\( \frac{B(\chi_{c2} \rightarrow \pi^+ \pi^-)}
        {B(\chi_{c2} \rightarrow K^+ K^-    )}
    = 1.88 \pm 0.51  \).

We thank the staffs of the BEPC Accelerator
and the Computing Center at the Institute of High Energy Physics,
Beijing, for their outstanding scientific efforts. This project
was partly supported by China Postdoctoral Science Foundation.
The work of the BES Collaboration was supported in part by 
the National Natural Science Foundation of China 
under Contract No. 19290400 and the Chinese Academy of Sciences 
under contract No. KJ85 (IHEP), and by the Department of
Energy under Contract Nos. DE-FG03-92ER40701 (Caltech), 
DE-FG03-93ER40788 (Colorado State University), DE-AC03-76SF00515 (SLAC), 
DE-FG03-91ER40679 (UC Irvine), DE-FG03-94ER40833 (U Hawaii), 
DE-FG03-95ER40925 (UT Dallas).

\begin{table}
\caption{Branching fractions of $\chi_{cJ} \rightarrow 0^-0^-$.
$B(\psi(2S) \rightarrow \gamma \chi_{c0}) = (9.3 \pm 0.8)\%$ and
$B(\psi(2S) \rightarrow \gamma \chi_{c2}) = (7.8 \pm 0.8)\%$ are 
used for branching fractions determination.
}
\begin{tabular}{|c|c|c||c|}
\hline
decay mode  & $N^{obs}$  & BR($\times 10^{-3}$)
	    & PDG~\cite{pdg}~($10^{-3}$) \\\hline
$\chi_{c0}\rightarrow \pi^+ \pi^-$ & $720 \pm 32$ & $4.68 \pm 0.26 \pm 0.65$
                                   & $7.5\pm 2.1$\\\hline
$\chi_{c0}\rightarrow K^+ K^-    $ & $774 \pm 38$ & $5.68 \pm 0.35 \pm 0.85$
                                   & $7.1\pm 2.4$\\\hline
$\chi_{c2}\rightarrow \pi^+ \pi^-$ & $185 \pm 16$ & $1.49 \pm 0.14 \pm 0.22$
                                   & $1.9\pm 1.0$\\\hline
$\chi_{c2}\rightarrow K^+ K^-    $ & $115 \pm 13$ & $0.79 \pm 0.14 \pm 0.13$
                                   & $1.5\pm 1.1$\\\hline
\end{tabular}
\label{pika}
\end{table}

\begin{table}
\caption{Branching ratios of $\chi_{cJ} \rightarrow p \bar{p}$.
$B(\psi(2S) \rightarrow \gamma \chi_{c0}) = (9.3 \pm 0.8)\%$,
$B(\psi(2S) \rightarrow \gamma \chi_{c1}) = (8.7 \pm 0.8)\%$ and
$B(\psi(2S) \rightarrow \gamma \chi_{c2}) = (7.8 \pm 0.8)\%$ are
used for branching fractions determination.
$\Gamma^{tot}_{\chi_{c0}}$ from this experiment,
$\Gamma^{tot}_{\chi_{c1}} = 0.88 \pm 0.14 \, \hbox{MeV}$ and
$\Gamma^{tot}_{\chi_{c2}} = 2.00 \pm 0.18 \, \hbox{MeV}$ are
used in calculating the partial widths.
}
\begin{tabular}{|c|c|c|c||c|c|}
\hline
     state   & $N^{obs}$ & BR($\times 10^{-5}$) & $\Gamma_{p\bar{p}}$ (keV)
	     & PDG~\cite{pdg}~BR($10^{-5}$)
	     & PDG~\cite{pdg}~$\Gamma_{p\bar{p}}$(keV)\\\hline
$\chi_{c0}$   & $15.2\pm 4.1$      & $15.9 \pm 4.3 \pm 5.3$
              & $2.3 \pm 1.1$      & $<90$  &  ---\\\hline
$\chi_{c1}$   & $ 4.2\pm 2.2$      & $4.2 \pm 2.2 \pm 2.8$
              & $0.037 \pm 0.032$  & $8.6\pm 1.2$ & $0.074 \pm 0.009$ \\\hline
$\chi_{c2}$   & $ 4.7\pm 2.5$      & $5.8 \pm 3.1 \pm 3.2$
              & $0.116 \pm 0.090$  & $10.0\pm 1.0$ & $0.206 \pm 0.022$ \\\hline
\end{tabular} 
\label{ppb}
\end{table}

\end{document}